\documentclass[sigconf]{acmart}
\acmConference[EASE 2026]{The 30th International Conference on Evaluation and Assessment in Software Engineering}{Tue 9 - Fri 12 June 2026}{Glasgow, United Kingdom}

\usepackage{fancyvrb}
\usepackage{xurl}
\usepackage{booktabs}
\usepackage{url}
\usepackage{amsmath,amsfonts}
\usepackage{algorithmic}
\usepackage{graphicx}
\usepackage{textcomp}
\usepackage{xspace}
\usepackage{subcaption}
\usepackage{tabularx}
\usepackage{framed}
\usepackage{chngpage}
\usepackage[many]{tcolorbox}
\usepackage{pgfplots}
\pgfplotsset{compat=newest}
\usepackage{siunitx}
\usepackage{balance}
\usepackage{enumitem}
\usepackage{listings} 
\usepackage{multirow} 
\usepackage{xfp}      

\usepackage{hyperref}

\usepackage{tikz}

\usepackage[group-separator={,}, group-minimum-digits=4]{siunitx}

\usepackage{threeparttable}

\newcommand{\rqa}{$RQ_1$}
\newcommand{\rqb}{$RQ_2$}
\newcommand{\rqc}{$RQ_3$}

\newcommand{\rqaa}{To what extent is AI-generated files maintained?}
\newcommand{\rqbb}{What types of maintenance activities are applied to AI-generated files?}
\newcommand{\rqcc}{Who maintains AI-generated files: AI agents or human developers?}

\newcommand{\rqA}{\rqa: \rqaa}
\newcommand{\rqB}{\rqb: \rqbb}
\newcommand{\rqC}{\rqc: \rqcc}

\newcommand{\TotalFuzzingProjectNum}{1311}
\newcommand{\TargetProjectNum}{628}

\pgfmathsetmacro{\TargetProjectRateNum}{round((\TargetProjectNum/\TotalFuzzingProjectNum)*100*10)/10}

\newcommand{\ExcludedProjectNum}{4}
\pgfmathsetmacro{\FinallyTargetProjectInt}{round(\TargetProjectNum - \ExcludedProjectNum)}

\newcommand{\AiGenerated}{AI-generated\xspace}
\newcommand{\HumanGenerated}{human-generated\xspace}

\definecolor{darkgreen}{rgb}{0, 0.5, 0} 
\definecolor{whitesmoke}{rgb}{0.99, 0.99, 0.99} 

\def\Underline{\setbox0\hbox\bgroup\let\\\endUnderline}
\def\endUnderline{\vphantom{y}\egroup\smash{\underline{\box0}}\\}
\def\|{\verb|}

\newcommand{\ie}{\textit{i.e.,}\xspace}
\newcommand{\eg}{\textit{e.g.,}\xspace}

\newcommand{\etal}{\xspace\textit{et al.}\xspace}

\newcommand{\motivation}{\noindent\textbf{Motivation. }}
\newcommand{\approach}{\smallskip\noindent\textbf{Approach. }}
\newcommand{\results}{\smallskip\noindent\textbf{Results. }}

\newcounter{findings_no}

\usepackage{listings}
\definecolor{backcolour}{rgb}{0.95,0.95,0.92}
\lstdefinelanguage{diff}{
  morecomment=**[f][\color{red}]{-},         
  morecomment=**[f][\color{darkgreen}]{+},       
  moredelim=**[is][\bfseries]{@@}{@@},
}
\definecolor{backcolour}{rgb}{0.95,0.95,0.92}
\lstdefinelanguage{commit}{ 
  breakindent = 0pt,
  numbers=none,
  backgroundcolor=\color{white},
  frame=single,
  xleftmargin=3.5em,
  numbersep=0em,
  xrightmargin=1.5em,
}

\usepackage[many]{tcolorbox}  
\tcbuselibrary{listings,breakable}
\definecolor{main}{HTML}{D0D3D4}    
\definecolor{sub}{HTML}{D0D3D4}     
\newtcolorbox{dbox}{
    left=2pt,right=2pt,top=2pt,bottom=2pt,
    enhanced, 
    boxrule = 0pt,
    enlarge top by=5pt,
    enlarge bottom by=3pt,
  }
\tcbset{
    sharp corners,
    before skip = 0.01cm,    
    after skip = 0.01cm      
}

\def\summarybox#1#2{
\medskip
\begin{tcolorbox}[
  enhanced,
  title=#1,
  colframe=darkgray,
  left=1mm,
  right=1mm,
  top=1mm,
  bottom=1mm,
]
    #2
\end{tcolorbox}
\medskip
}

\AtBeginDocument{%
  \providecommand\BibTeX{{%
    \normalfont B\kern-0.5em{\scshape i\kern-0.25em b}\kern-0.8em\TeX}}}

\setcopyright{acmlicensed}
\copyrightyear{2025}
\acmYear{2025}
\acmDOI{XXXXXXX.XXXXXXX}

\begin{document}

\title{To What Extent Does Agent-generated Code Require Maintenance? An Empirical Study}

\author{Shota Sawada}
\affiliation{%
  \institution{National Institute of Technology, Nara College}
  \city{Yamatokoriyama}
  \country{Japan}
}
\email{AI1218@nara.kosen-ac.jp}

\author{Tatsuya Shirai}
\affiliation{%
  \institution{Nara Institute of Science and Technology}
  \city{Ikoma}
  \country{Japan}
}
\email{shirai.tatsuya.sp1@naist.ac.jp}

\author{Yutaro Kashiwa}
\affiliation{%
  \institution{Nara Institute of Science and Technology}
  \city{Ikoma}
  \country{Japan}
}
\email{yutaro.kashiwa@is.naist.jp}

\author{Ken’ichi Yamaguchi}
\affiliation{%
  \institution{National Institute of Technology, Nara College}
  \city{Yamatokoriyama}
  \country{Japan}
}
\email{yamaguti@info.nara-k.ac.jp}

\author{Hiroshi Iwata}
\affiliation{%
  \institution{National Institute of Technology, Nara College}
  \city{Yamatokoriyama}
  \country{Japan}
}
\email{iwata@info.nara-k.ac.jp}

\author{Hajimu Iida}
\affiliation{%
  \institution{Nara Institute of Science and Technology}
  \city{Ikoma}
  \country{Japan}
}
\email{iida@itc.naist.jp}

\renewcommand{\shortauthors}{Sawada, et al.}

\begin{abstract}

LLM-based autonomous coding agents have reshaped software development. While these agents excel at code generation, open questions persist about the long-term maintainability of AI-generated code. This study empirically investigates the maintenance extent, human involvement, and modification types of AI-generated files versus human-authored code.
Using the AIDev dataset of AI-generated pull requests and GitHub, we analyzed over 1,000 files and approximately 3,200 changes from 100 popular repositories.

Our findings show that:
(i) AI-generated files receive less frequent maintenance than human-authored code, with updates affecting only a small fraction of file size;
(ii) the most frequent modifications to AI code are feature extensions, whereas human updates focus on bug fixes, and (iii) human developers perform the large majority of this maintenance.

\end{abstract}



\maketitle

\section{Introduction}\label{sec:introduction}

Software maintenance is critical in the software development lifecycle to keep systems reliable, stable, and compatible with evolving technology. Developers spend considerable time on maintenance activities, including bug fixing, refactoring, and feature extensions. Previous studies~\cite{schach2007object, Ogheneovo2014OnTR} have investigated software maintenance to understand and reduce these costs. Dehaghani\etal\cite{dehaghani2013factors} report that approximately 90\% of the software development life cycle is related to maintenance activities, and that these costs have increased by 50\% over the past two decades.

Most of these studies, however, were conducted before the emergence of large language models (LLMs). In particular, Agentic Coding has transformed software development. Agentic Coding is an approach in which AI agents autonomously decompose high-level instructions into subtasks and perform coding activities such as writing, debugging, and refactoring with minimal human intervention. While Agentic Coding accelerates development, prior studies indicate it introduces potential risks to code quality~\cite{DBLP:journals/corr/abs-2510-03029, DBLP:journals/corr/abs-2504-12608, DBLP:journals/corr/abs-2508-21634}. He\etal~\cite{he2025doesaiassistedcodingdeliver} revealed that although agentic coding produces a sharp immediate increase in development velocity, it results in a 30\% rise in static analysis warnings and a 41\% rise in code complexity over the long term. Sankhe\etal\cite{Sankhe2025EmpiricalAO} confirmed that while AI assistance improved productivity by 31.4\%, it also led to a 23.7\% increase in security vulnerabilities and a notable rise in code duplication.

These agent-related studies have primarily focused on short-term effects, such as before and after the change. The maintenance activities required after \AiGenerated files are introduced have received little attention, and the extent to which human intervention is required is still unclear. This study empirically investigates the maintainability of \AiGenerated files and the participation of AI in maintenance activities within projects that adopt autonomous coding agents. Specifically, we identify files and commits added by four major AI agents (Copilot, Claude, Devin, and Cursor) from the repositories and PR history recorded in the AIDev dataset~\cite{DBLP:journals/corr/abs-2507-15003}, and compare them with \HumanGenerated files and commits to clarify the maintenance activities required.

Our empirical analysis of 3,238 commits shows that \AiGenerated files receive significantly less maintenance than human files, yet human developers perform about 83\% of it. Modifications to \AiGenerated files are mostly feature extensions, while \HumanGenerated files focus on bug fixes. 

\smallskip\noindent\textbf{Replication Packages:} To facilitate replication and further
studies, we provide the scripts and data used in our replication package.\footnote{\url{https://github.com/ShouSawa/AI-Code-Maintainability}}

\begin{figure*}[!t]
    \centering
    \includegraphics[width=0.91\linewidth]{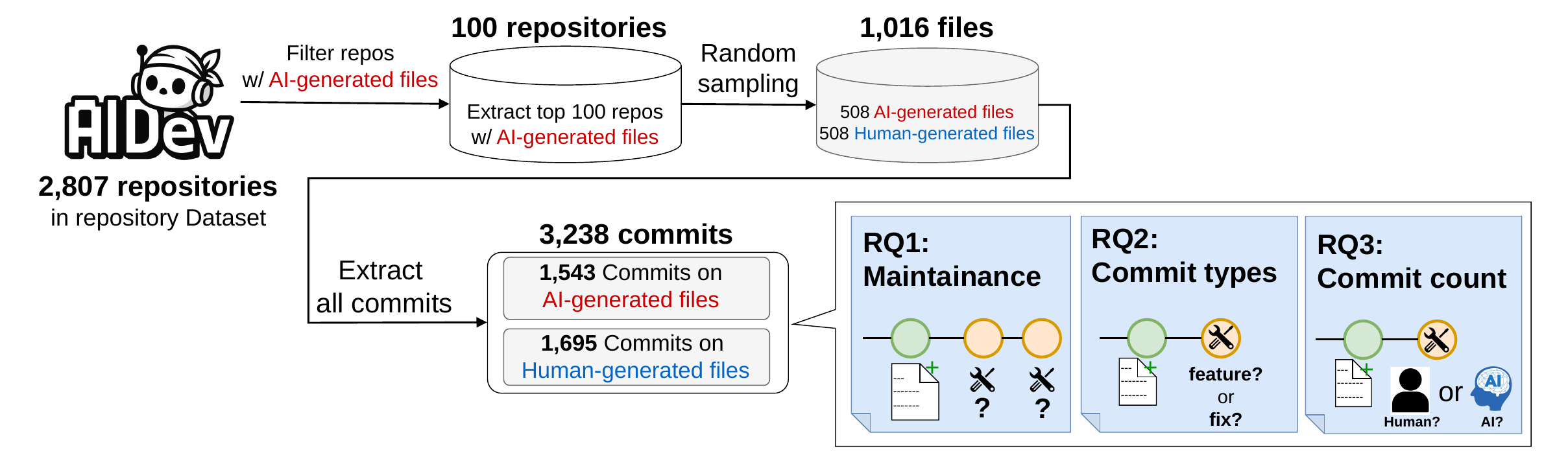}
    \caption{Overview of our data collection process and research questions}
    \label{fig:dataflow}
\end{figure*}

\section{Related Work}\label{sec:background}

While Agentic Coding accelerates code implementation~\cite{DBLP:conf/iclr/0001LSXTZPSLSTL25, DBLP:conf/iclr/HongZCZCWZWYLZR24}, a growing body of work~\cite{DBLP:conf/chi/Vaithilingam0G22, DBLP:journals/corr/abs-2405-16081} has begun to examine its broader effects on development tasks. Cihan\etal\cite{DBLP:conf/icse-seip/CihanHIGDBUT25} reported that introducing automated code review tools increased the average time required to close pull requests. Becker\etal\cite{DBLP:journals/corr/abs-2507-09089} conducted a controlled experiment and found that AI tool usage led to a 19\% increase in task completion time.

The quality of generated code has drawn considerable attention~\cite{DBLP:journals/corr/abs-2501-16857}. Paul\etal\cite{DBLP:journals/corr/abs-2510-03029} compared code smells in human-written and \AiGenerated code, finding that \AiGenerated code contains 63\% more code smells on average. He\etal\cite{he2025doesaiassistedcodingdeliver} reported that although development speed improves temporarily, technical debt accumulates through static analysis warnings and code complexity, leading to a decline in future velocity. From a security perspective, several studies indicate that \AiGenerated code is more prone to vulnerabilities than human-written code~\cite{DBLP:conf/ccs/PerryS0B23, DBLP:journals/corr/abs-2508-21634}. Pearce\etal\cite{DBLP:journals/cacm/PearceATDK25} evaluated code generated by GitHub Copilot against security scenarios from MITRE's ``CWE Top 25''\footnote{\url{https://cwe.mitre.org/top25/}} and found that 40\% of the generated programs contained vulnerabilities.

Maintainability issues, such as modularity and readability, have also been identified~\cite{DBLP:conf/saner/NunesFSNFS25}. Liu\etal\cite{DBLP:journals/corr/abs-2504-12608} investigated repetition in \AiGenerated code and revealed prevalent redundant repetitions at both the character and block levels, which degrade readability and efficiency. Kravchuk-Kirilyuk\etal\cite{KravchukKirilyuk2025} reported that while \AiGenerated code can mimic superficial modular structures, it tends to violate principles that sustain maintainability, such as encapsulation, and introduces hidden dependencies. Watanabe\etal\cite{watanabe2026cutpredictingunnecessarymethods} analyzed agent-generated pull requests and found that 9.9\% of the generated methods are eventually deleted during review, placing an unnecessary cognitive burden on human reviewers.

Beyond generation-time quality, the maintenance activities surrounding \AiGenerated code present additional challenges. Haque\etal\cite{DBLP:journals/corr/abs-2601-03556} reported that test code included in initial PRs authored by AI agents is often insufficient and frequently requires additional updates after the initial PR. Ottenhof\etal\cite{DBLP:journals/corr/abs-2601-20160} conducted an empirical study on agentic refactoring and found that while human developers perform diverse structural improvements, refactorings by AI agents are dominated by superficial changes such as adding or modifying annotations. AI agents also struggle with continuous revisions during review. Minh\etal\cite{DBLP:journals/corr/abs-2601-00753} analyzed agent-authored PRs and found that although agents excel at narrow automation, they frequently fail at iterative refinement, leading to ``ghosting'' (abandonment of PRs) when faced with subjective human feedback.

These prior studies primarily evaluate code quality at the point of generation or examine short-term effects immediately after AI tool adoption. However, once \AiGenerated files are merged into a codebase, human developers must continue to maintain them over extended periods. This long-term maintenance burden has received little empirical investigation. Our study addresses this gap by tracking the maintenance activities that follow the introduction of \AiGenerated files and quantifying the extent of human intervention required to sustain them.

\section{Data Collection}\label{sec:studydesign}

To analyze the maintenance of \AiGenerated code, we need to identify files created by AI agents and track their subsequent commit histories. We used the AIDev dataset~\cite{DBLP:journals/corr/abs-2507-15003}, which contains more than 456,000 pull requests made by five autonomous coding agents (OpenAI Codex, Devin, GitHub Copilot, Cursor, and Claude Code) from approximately 61,000 repositories. All PRs in this dataset were created between December 2024 and July 2025, following the release of agentic coding tools.

To ensure the quality of the analyzed projects, we utilized the repository list provided alongside the AIDev dataset, which consists of 2,807 repositories that have already been filtered to include only those with more than 100 stars. From this repository dataset, we identified AI-generated files through the following steps. 
First, we extracted files created in these PRs by identifying files marked as ``added'' in the Git file status of each PR. Second, we verified that each file was created by an AI agent by examining the committer name. Although the dataset guarantees that PRs are created by agents, individual commits may be made by human developers.
We matched committer names against AI-agent account identifiers (\ie bots): ``claude[bot]''\footnote{\url{https://github.com/claude}} for Claude Code, ``Cursor Agent''\footnote{\url{https://github.com/apps/cursor}} for Cursor, ``Copilot''\footnote{\url{https://docs.github.com/en/enterprise-cloud@latest/copilot/concepts/agents/coding-agent/about-coding-agent}} for GitHub Copilot, and ``devin-ai-integration''\footnote{\url{https://github.com/apps/devin-ai-integration}} for Devin. We excluded Codex PRs because it does not appear as a commit owner, making it difficult to determine whether a commit was created by an agent or a developer~\cite{0_Codex_Onwer}.
Using this approach, we collected files generated by agents from 100 repositories.
We restricted the sample to 100 repositories to reduce the high computational cost and address API limits.

Due to a large number of files from specific projects, we randomly sampled up to ten \AiGenerated files from each repository. For comparison, we also sampled up to ten \HumanGenerated files from the same repository created during the same period. When fewer than ten files were available in either category, we extracted all available files and balanced the dataset by selecting an equal number from the other category. For each selected file, we collected all commits recorded through January 31, 2026. This ensures a maintenance observation period of at least six months for all files because we obtained files created before July 31.

In total, we collected 508 \AiGenerated files and 508 \HumanGenerated files from 100 repositories, along with 1,543 commits modifying \AiGenerated files and 1,695 commits modifying \HumanGenerated files.
We excluded the initial file-creation commits from this analysis, as we focus on subsequent maintenance activities.


\section{Research Questions}\label{sec:results}
\subsection*{\rqA}\label{sec:rqa}
\motivation
Prior studies have indicated that \AiGenerated code tends to be lower quality compared to human-written code~\cite{DBLP:journals/corr/abs-2510-03029, DBLP:journals/corr/abs-2508-21634, he2025doesaiassistedcodingdeliver}. If \AiGenerated code demands significantly more maintenance, the productivity benefits of using AI agents may be offset by long-term maintenance costs. However, little is known about how \AiGenerated code is actually maintained after its initial creation.

\approach
In this study, we analyzed the extent of maintenance of \AiGenerated files from two perspectives: the maintenance frequency and the magnitude of the maintenance.

For the maintenance frequency, we compared the number of commits applied to \AiGenerated files with those applied to \HumanGenerated files to evaluate the maintenance activity associated with \AiGenerated code.
Specifically, we aggregated commits for each file and grouped them into one-month intervals relative to the creation date. We then focused on the first six months, a period common to all files, and visualized the distributions using violin plots. We use only six months because the observation periods after file creation vary across files, making direct comparison beyond this window unreliable. Note that as we described in the Data Collection section, we ensured that at least six months have passed since the creation of all the studied files. Additionally, we did not include the first commit that created the file as maintenance.

For the magnitude, we examined whether there were differences in the number of lines of code changed per commit. However, the raw number of changed lines depends on file size, and larger files tend to have more lines modified. We normalize the changes by the size of the files to measure the percentage of changes.

\begin{figure*}[t]
    \centering
    \begin{subfigure}{0.49\textwidth}
        \centering
        \includegraphics[width=1\linewidth]{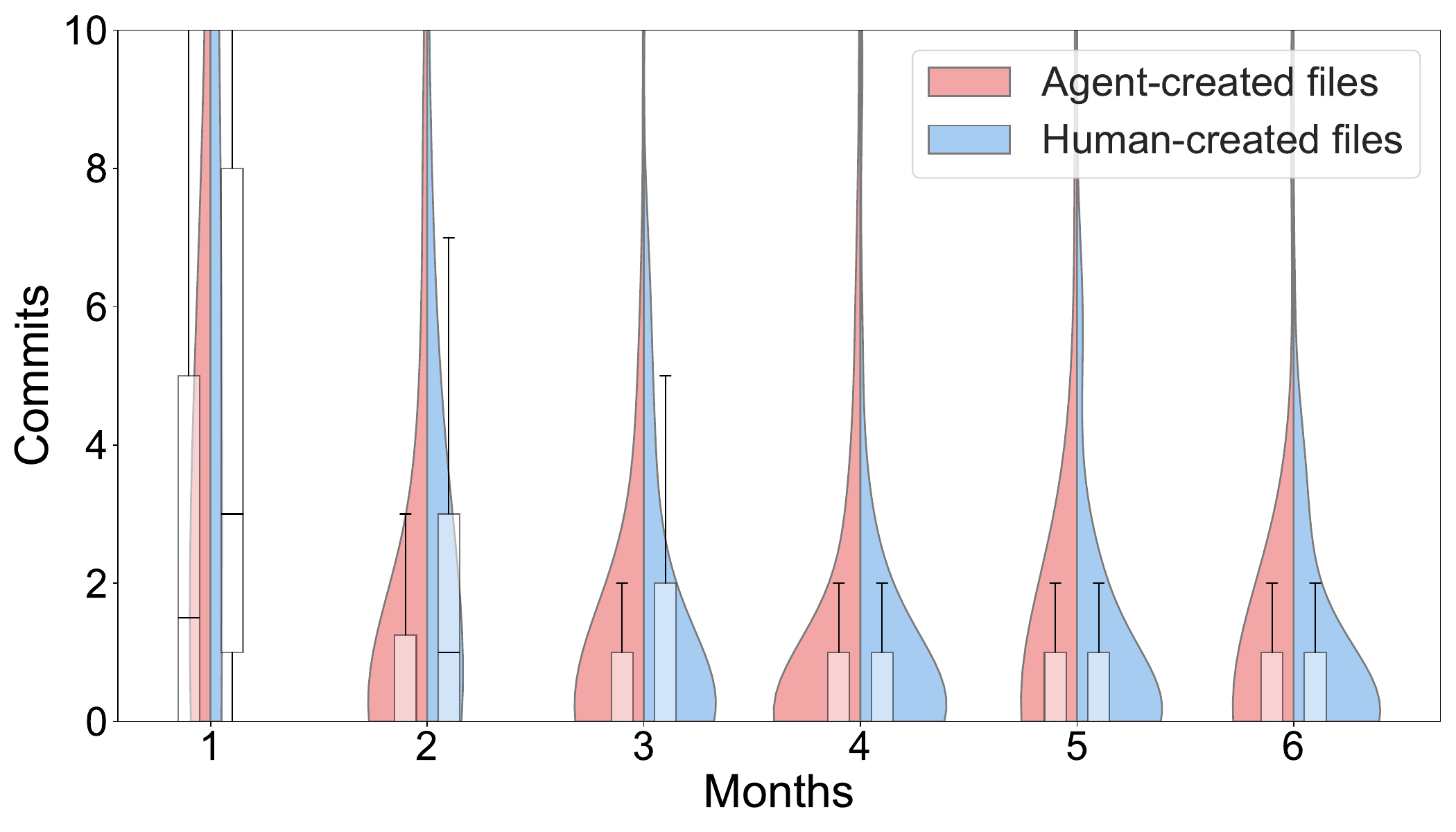}
        \caption{Number of commits per month after file creation}
        \label{fig:RQ1_count}
    \end{subfigure}
    \hfill
    \begin{subfigure}{0.49\textwidth}
        \centering
        \includegraphics[width=1\linewidth]{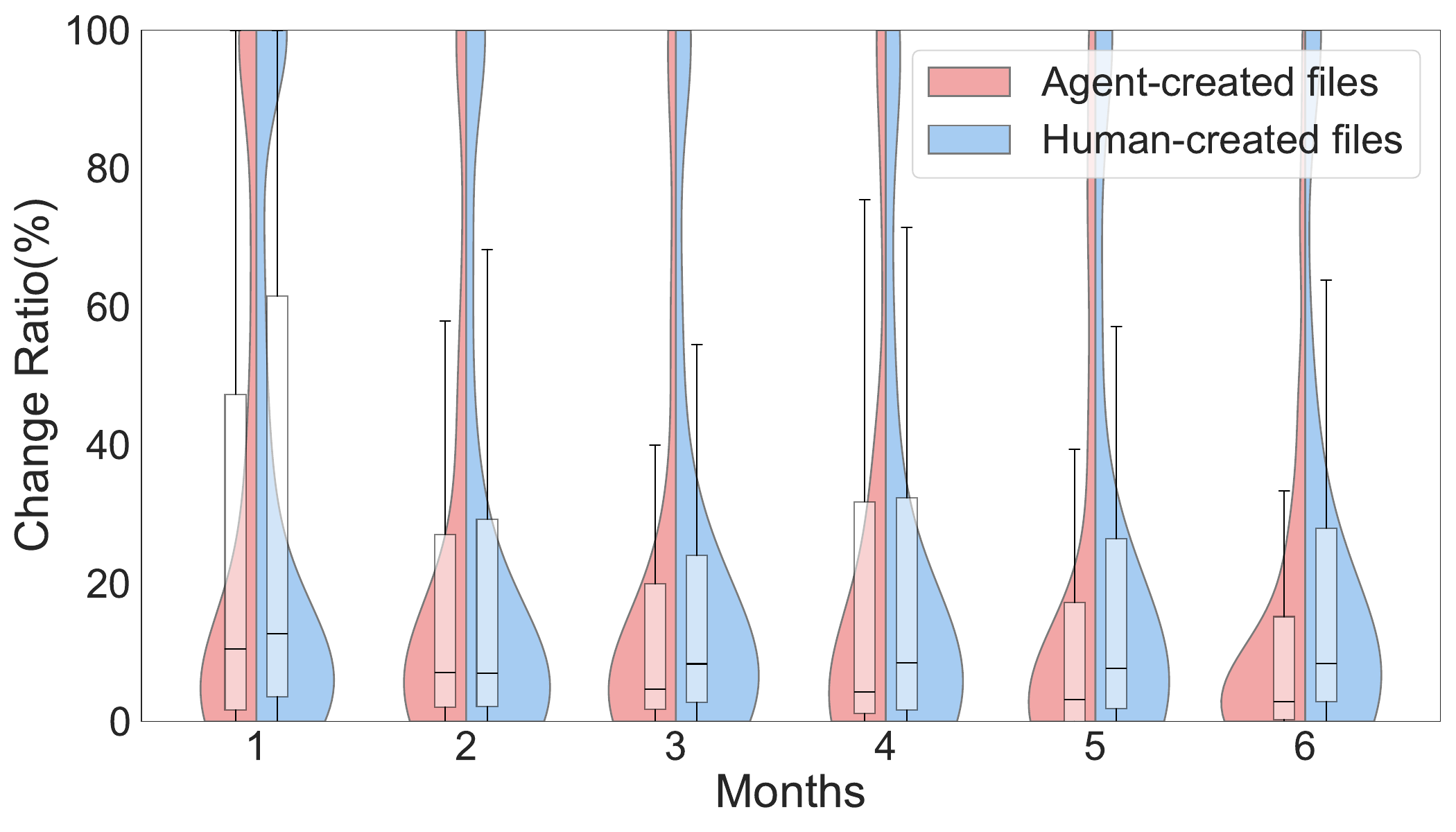}
        \caption{Percentage of lines changed per month after file creation}
        \label{fig:RQ1_change_ratio}
    \end{subfigure}
    \caption{The maintenance frequency and magnitude for Agent and Human generated files}
    \label{fig:RQ1_combined}
    \vspace{7mm}
\end{figure*}

\results
\autoref{fig:RQ1_count} shows the number of commits for each file. The red plots represent \AiGenerated files, while the blue plots represent \HumanGenerated files. 
The horizontal axis indicates the time elapsed since file creation. 
To mitigate the influence of extreme values, outliers are excluded from the visualizations.

\AiGenerated files exhibited approximately half the commit count of \HumanGenerated files during the first month, with maintenance activity gradually declining over the subsequent three months. 
The distribution shows that maintenance continues at a reduced frequency rather than stopping entirely. 
These findings imply that \AiGenerated code does not impose an immediate, severe burden on developers, and that more maintenance effort is currently devoted to \HumanGenerated code.

From the fourth month onward, the gap with \HumanGenerated files narrows. 
The commit count stays small but non-zero, indicating that maintenance continues over the long term to a similar extent as for \HumanGenerated files. 
In summary, although the need for maintenance immediately after generation is lower for \AiGenerated files, ongoing maintenance is still required.

Next, \autoref{fig:RQ1_change_ratio} shows the distribution of the magnitude of maintenance for each month after file creation. 
The results reveal that \HumanGenerated files exhibit a larger magnitude compared to \AiGenerated files. 
The smaller magnitude of changes in \AiGenerated files implies that maintenance is often limited to minor changes or slight modifications. 
Furthermore, a decreasing trend was observed in the modification ratio of \AiGenerated files. 
This indicates that the volume of maintenance decreases over time. 
In contrast, the higher ratio in \HumanGenerated files likely reflects more fundamental structural changes and active refactoring. 
Ultimately, \AiGenerated code does not impose an immediate, severe maintenance burden on developers. 
Moreover, although maintenance is required in the long term, its magnitude is smaller and less burdensome compared to \HumanGenerated files.

\summarybox{\textbf{Answer to RQ1.}}{
AI-generated files require less maintenance than human-generated files in both frequency and magnitude. This suggests that AI-generated code does not impose a significant maintenance burden on developers.
}

\subsection*{\rqB}\label{sec:rqb}
\motivation
Watanabe\etal\cite{DBLP:journals/corr/abs-2509-14745} show that agents perform bug fixing more frequently than feature addition. However, their study focuses only on commits within PRs generated by AI agents. It remains unclear what types of changes are made in subsequent maintenance commits after file creation. In this RQ, we examine whether modifications to \AiGenerated files are primarily driven by constructive activities (\eg feature additions) or by corrective actions (\eg bug fixes or refactoring).

\approach
We categorized commits modifying \AiGenerated files and \HumanGenerated files using the Conventional Commits Classification System (CCS)~\cite{DBLP:conf/icse/ZengZQL25}, an automatic classification system that defines ten commit types. CCS extends the original Conventional Commits specification\footnote{\url{https://www.conventionalcommits.org/en/v1.0.0/}} by providing clearer, non-overlapping definitions. This system achieves 77.95\% precision and has been widely adopted in software engineering research~\cite{DBLP:journals/corr/abs-2509-14745}.

\begin{table*}[!t]
    \centering
    \begin{threeparttable}
        \caption{Types of maintenance activities for Human-generated files and AI-generated files}
        \label{tab:summary_ccs}
        \begin{tabular}{lp{7cm}rrr}
            \toprule
            \textbf{Category} & \textbf{Description} & \textbf{Human (\%)} & \textbf{AI (\%)} & \textbf{Diff (\%)}\\
            \midrule
            feat     & Addition of new features                       & 256 (15.10\%) & \textbf{336 (21.78\%)} & \textbf{+6.68} \\
            build    & Changes related to the build system           & 136 (\phantom{0}8.02\%)  & 155 (10.05\%) & +2.03 \\
            style    & Code style changes                            & \phantom{0}67 (\phantom{0}3.95\%)  & \phantom{0}75 (\phantom{0}4.86\%)  & +0.91  \\
            chore    & Miscellaneous tasks or other changes          & 195 (11.50\%) & 206 (13.35\%) & +1.85 \\
            perf     & Performance improvements                      & \phantom{0}38 (\phantom{0}2.24\%)  & \phantom{0}31 (\phantom{0}2.01\%)  & -0.23  \\
            ci       & Changes related to continuous integration (CI)& \phantom{0}77 (\phantom{0}4.54\%)  & \phantom{0}69 (\phantom{0}4.47\%)  & -0.07   \\
            test     & Addition or improvement of tests              & 107 (\phantom{0}6.31\%)  & \phantom{0}94 (\phantom{0}6.09\%)  & -0.22  \\
            refactor & Code refactoring (internal structural improvements) & 256 (15.10\%) & 219 (14.19\%) & -0.91 \\
            docs     & Documentation-only changes                    & 275 (16.22\%) & 166 (10.76\%) & -5.46 \\
            fix      & Bug fixes                                     & \textbf{284 (16.76\%)} & 181 (11.73\%) & \textbf{-5.03} \\
            \bottomrule
        \end{tabular}
        \begin{tablenotes}[flushleft]
            \footnotesize
            \item \textit{Note:} Total commits: Human = 1,695, AI = 1,543. Percentages are calculated based on these totals. Bold numbers indicate the highest frequency in each column and the largest percentage differences. Minor categories (e.g., revert, i18n) with $<1\%$ frequency are omitted for brevity.
        \end{tablenotes}
    \end{threeparttable}
\end{table*}

\smallskip
\results
\autoref{tab:summary_ccs} summarizes the distribution of commit types for \AiGenerated files and \HumanGenerated files.
For \AiGenerated files, the most frequent commit category was feat (21.78\%), followed by refactor (14.19\%), chore (13.35\%), and fix (11.73\%).
In contrast, for \HumanGenerated files, fix was the most frequent category (16.76\%), followed by docs (16.22\%), refactor (15.10\%), and feat (15.10\%).

\smallskip
Compared to \HumanGenerated files, \AiGenerated files exhibited a lower proportion of maintenance-oriented commits such as fix and refactor, indicating fewer corrective or rework-related modifications to the generated code.
Meanwhile, \AiGenerated files showed a notably higher proportion of feat commits, suggesting that many changes were related to functional extensions rather than corrections.
In addition, fix and docs commits were more prevalent in \HumanGenerated files than in \AiGenerated files.

\smallskip
These results suggest that while \AiGenerated code does not frequently require bug-fixing changes, it often lacks sufficient coverage of requirements or generality, thereby necessitating additional feature implementations after the initial generation.
For less frequent categories such as test, ci, and style, no substantial differences were observed between \AiGenerated and \HumanGenerated files, with both exhibiting similar proportions.


\bigskip
\summarybox{\textbf{Answer to RQ2.}}{
Maintenance of \AiGenerated files consists mainly of feature-related changes, whereas corrective and documentation-related changes are less common than in \HumanGenerated files.
}

\subsection*{\rqC}\label{sec:rqc}

\motivation
As AI coding agents become more capable, they are now used not only for initial code generation but also for ongoing tasks such as bug fixes and feature additions \cite{he2025doesaiassistedcodingdeliver,DBLP:journals/corr/abs-2511-04824}. This raises an important question: \textit{once AI agents create files, who is responsible for maintaining them?}
If human developers bear this burden, they must understand and modify code they did not write. Despite the growing use of AI agents, there is limited empirical understanding of who actually performs maintenance after file creation. 


\approach
For each commit modifying the studied files, we identified whether the author was an AI bot account or a human developer by matching committer names against known AI-agent account identifiers. We then calculated the proportion of commits made by AI agents and by human developers for both \AiGenerated files and \HumanGenerated files.

\begin{table}[t]
    \centering
    \caption{Commits to AI and human-generated files}
    \begin{tabular}{lrrr}
        \toprule
        \textbf{Category} & \textbf{AI Commits} & \textbf{Human Commits} & \textbf{Total} \\
        \midrule
        AI files & 259 (16.79\%) & 1284 (83.21\%) & 1543\\
        Human files & 119 (7.02\%) & 1576 (92.98\%) & 1695\\
        \bottomrule
    \end{tabular}
    \label{tab:RQ2_table}
\end{table}
\results
We extracted 1,543 commits modifying \AiGenerated files and 1,695 commits modifying \HumanGenerated files. For \AiGenerated files, 83.21\% of commits (1,284) were made by human developers, while only 16.79\% (259) were made by AI agents. Similarly, for \HumanGenerated files, human developers performed 92.98\% of commits (1,576), with AI agents contributing only 7.02\% (119). These results indicate that maintenance is the most frequently performed by humans regardless of whether the file was originally created by an AI agent or a human developer. Notably, AI agents are more frequently involved in maintaining \AiGenerated files (16.79\%) than \HumanGenerated files (7.02\%), but human developers still handle the large majority of maintenance in both cases.

\summarybox{\textbf{Answer to RQ3}}{
Human developers perform the majority of maintenance for \AiGenerated files, while AI agents account for only a small portion of maintenance activity.
}

\section{Implication}\label{sec:implications}
This section discusses the implications of our findings for researchers and practitioners.

\vspace{1mm}\noindent\textbf{\textit{Practitioners should monitor AI-generated code beyond initial integration.}}
AI-generated files receive less frequent maintenance and smaller magnitudes of change compared to human-authored code (RQ1). While this may suggest that AI-generated code is stable enough to function without immediate corrective action, alternative explanations are also possible; for example, developers may avoid modifying AI-generated code due to difficulty in comprehending it, or some generated files may not be actively exercised in production. Practitioners should therefore not assume that low maintenance frequency alone indicates high code quality. 
Moreover, subsequent maintenance predominantly involves feature extensions rather than bug fixes (RQ2), with humans still performing approximately 83\% of these modifications (RQ3).
This pattern indicates that even when AI-generated code operates without critical defects, it may lack full requirement coverage, necessitating human developers to build upon the initial generation. Teams adopting AI agents should establish review processes that track how generated files evolve after merging, rather than treating the initial generation as a finished product.

\vspace{1mm}\noindent\textbf{\textit{Researchers should develop AI agents for long-term maintenance.}}
AI agents currently account for only about 17\% of maintenance activity on the files they create. Agents are effective at initial generation but contribute minimally to the sustained maintenance lifecycle. Future research should prioritize creating agents capable of performing complex refactoring and functional growth. Investigating why AI-generated code requires more functional additions could lead to agents that better anticipate future requirements, improving the long-term utility of AI-generated software.

\section{Future Direction}

\textbf{Extend the observation period.} Our analysis covers only six months after file creation because agentic coding tools are relatively new and sufficient data points were not yet available. As these tools mature and accumulate longer histories, extending the observation window will allow us to examine whether maintenance patterns stabilize, shift, or accelerate over time. RQ1 revealed that the maintenance gap between AI-generated and human-generated files narrows from the fourth month onward, raising the question of whether AI-generated code eventually converges to or exceeds the maintenance burden of human-authored code. A longer observation period would also enable comparison across different AI agents, as each may exhibit distinct maintenance trajectories depending on its code generation strategy and the projects it is applied to.

\smallskip\noindent\textbf{Predict maintenance risk from code characteristics.}  Our current analysis focuses on the frequency and magnitude of maintenance, but does not examine what properties of the generated code correlate with higher maintenance needs. Future work should investigate whether code-level features such as cyclomatic complexity, coupling, and file size can serve as predictors of subsequent maintenance activity. For instance, identifying which combinations of these features distinguish files that stabilize quickly from those that demand sustained attention would yield actionable insights. Such a predictive model could serve as a practical tool for development teams to prioritize review effort on AI-generated files that are most likely to require future intervention.

\smallskip\noindent\textbf{Develop AI agents specialized in code maintenance.} 
As shown in RQ2, the most common maintenance activity is feature extension, which requires understanding evolving project requirements and broader codebase context, tasks that go well beyond isolated code generation. Moreover, RQ3 reveals that AI agents contribute only approximately 17\% of maintenance on the files they create, indicating a clear gap in their current capabilities.
These findings motivate the development of maintenance-oriented agents capable of tracking requirement changes and performing targeted modifications on existing code. Combined with the risk prediction approach described above, such agents could form an end-to-end pipeline that first identifies files likely to need maintenance and then applies appropriate modifications automatically.

\section{Threats to Validity}\label{sec:threats_to_validity}

\noindent\textbf{Internal validity:}
We categorized maintenance activities by commit type but did not control for functional complexity, which may differ between \AiGenerated and \HumanGenerated files. We mitigated this by randomly sampling an equal number of files from the same repositories per category, though some bias may persist. File importance and intended usage are also uncontrolled: developers may avoid AI for critical files or apply it only in specific cases.

\noindent\textbf{Construct validity:}
The dataset covers the first six months after agentic coding tools were released, when developers are likely early adopters prioritizing feature development over maintenance. We also cannot detect AI-assisted code undeclared in commit messages, so \HumanGenerated files may include such code.

\noindent\textbf{External validity:}
We studied only 508 \AiGenerated files from 100 repositories, which limits generalizability. We also excluded Codex despite its many commits. Future work should expand the dataset to address these limitations.
\section{Conclusions}\label{sec:relatedwork}
This empirical study examined the long-term maintainability of code produced by autonomous coding agents. Our results show that \AiGenerated files require less frequent modifications than \HumanGenerated files, and these modifications tend to be minor, affecting a smaller proportion of file size, with activity declining after the first month. In contrast, \HumanGenerated code receives more active structural changes throughout the observation period.

The nature of maintenance also differs: updates to \AiGenerated files are dominated by feature extensions (22\%), whereas updates to \HumanGenerated files focus on bug fixes and documentation. Human developers perform approximately 83\% of maintenance on \AiGenerated files, with AI agents contributing only 17\%. Although the lower maintenance frequency may suggest stable code, alternative explanations such as limited usage or difficulty in comprehension cannot be ruled out.

Future research should extend the observation period beyond six months, compare maintenance patterns across different AI agents, investigate which code characteristics correlate with higher maintenance needs, and explore the development of maintenance-oriented agents that can reduce the human burden identified in this study.

\begin{acks}
We gratefully acknowledge the financial support of JSPS KAKENHI grants (JP24K02921, JP25K03100), as well as JST PRESTO grant (JPMJPR22P3), ASPIRE grant (JPMJAP2415), and CREST grant (JPMJCR23M1, JPMJCR26X7).

\end{acks}

\balance
\bibliographystyle{unsrt}
\bibliography{references}


\end{document}